\documentclass[12pt]{amsart}
\usepackage{amsmath, amssymb, amsthm}
\usepackage[width=6.5in, height=8in]{geometry}
\usepackage{setspace}
\newtheorem{theorem}{Theorem}[section]
\newtheorem{lemma}[theorem]{Lemma}
\newtheorem{proposition}[theorem]{Proposition}
\newtheorem{corollary}[theorem]{Corollary} \theoremstyle{definition}
 \theoremstyle{remark}

\newcommand{\C}{\mathbb C}

\newcommand{\N}{\mathbb N}
\newcommand{\Z}{\mathbb Z}
\newcommand{\R}{\mathbb R}
\renewcommand{\P}{\mathcal P}
\newcommand{\Iscr}{\mathcal I}
\newcommand{\Bscr}{\mathcal B}
\newcommand{\Lscr}{\mathcal L}
\newcommand{\Wscr}{\mathcal W}

\newcommand{\hf}{\frac{1}{2}}

\newcommand{\m}{\mu}
\newcommand{\V}{\|}

\newcommand{\G}{\Gamma}
\renewcommand{\t}{\tau}
\renewcommand{\b}{\beta}

\newcommand{\sq}{\square}
\newcommand{\z}{\zeta}
\newcommand{\D}{\Delta}
\renewcommand{\a}{\alpha}
\renewcommand{\l}{\lambda}
\newcommand{\g}{\gamma}
\renewcommand{\d}{\delta}

\newcommand{\Res}{\text{Res }}
\renewcommand{\and}{\text{~~ and ~~}}
\renewcommand{\part}{\partial}
\newcommand{\ra}{\rightarrow}

\begin{document}

\title [Meromorphic Solutions to a  Differential--Difference Equation]
{Meromorphic Solutions to a  Differential--Difference Equation Describing\\
        Certain Self-Similar Potentials}
\author{ Alexander Tovbis}

\begin{abstract}
    In this paper we prove the existence of meromorphic solutions to a 
    nonlinear differential difference equation that describe certain self-similar
    potentials for the Schroedinger operator.
\end{abstract}

\maketitle
\maketitle

\section{Introduction}
\label{sec:introduction}

Let $L=-\part^2_x+u(x)$ be a Schrodinger operator that we factorize
as $L=A^+A^- +\l$, where $A^\pm=\mp\part_x+f(x)$ and $x,\l\in\C$. Then the function
$f(x)$ satisfies the Riccati equation $f^2(x)-f'(x)+\l=u(x)$. If $\tilde L=A^-A^+ +\l$ 
denotes a new Schroedinger operator, obtained from $L$ by permuting the operator factors
$A^\pm$, then the  potential $\tilde u(x)$ of $\tilde L$ is given by 
$\tilde u(x)=f^2(x)+f'(x)+\l$.
The differential-difference equation (DDE)
\begin{equation}
    \label{eq:1i}
    [f(x)+f(x+a)]'+f^2(x)-f^2(x+a)=\m,
\end{equation}
where $a$ and $\m$ are some complex constants, was derived in \cite{S1,S2}
to describe potentials of the Schrodinger operator $L$ satisfying the self-similarity
constraint $\tilde L=ULU^{-1}+\m$, where $U$ is the translation operator
$Uf(x)=f(x+a)$. In the case $\m=0$ equation \eqref{eq:1i} is satisfied 
by  ( \cite {S1,S2})
\begin{equation}
    \label{eq:2i}
    f(x)=-\hf \frac {\P'(x-x_0)-\P'(a)}{\P(x-x_0)-\P(a)},
\end{equation}
where $\P$ denotes the Weierstrass elliptic function and $x_0\in \C$ is
an arbitrary constant. The function \eqref{eq:2i} is a meromorphic function 
with only simple poles.

The aim of the present paper is to prove existence of meromorphic solutions 
with simple poles to \eqref{eq:1i} in the case $\m\not =0$. 
Without any loss
of generality, we assume $a$ to be a positive real number. Indeed, if 
$a=re^{i\phi}$, where $\phi\in\R$ and $r>0$, then the transformation
$x\ra e^{i\phi}x$, $f\ra e^{-i\phi}f$ reduces \eqref{eq:1i}
to an equation of the same type with the  step $r\in\R^+$.
Our approach, which utilizes some ideas of \cite {To}, 
consists of essentially three statements: 1) there exist two different formal
power series solutions (in powers of $x^{-\hf}$) to \eqref{eq:1i}; 2) for any
formal solution $\hat f(x)$ there exists an actual solution $f(x)$, analytic
(when $|x|$ is sufficiently large)
in a sector $S$ on the complex $x$-plane of opening greater than $\pi$ and having
the asymptotic expansion $\hat f(x)$ in $S$; 3) any such solution can be meromorphically 
continued onto $\C$ so that $f(x)$ may have only first order poles.
 These  statements are proven in Sections
 \ref{sec:asymp} and \ref{sec:mer} respectively.   The author wants express his gratitude
to  V. Spiridonov for interesting discussions held at the 
NATO ASI ``Special functions 2000", Tempe, Arizona and for the following correspondence.
The author also want to use this opportunity to thank the organizers of the 
NATO ASI ``Special functions 2000".

\section{Meromorphic continuation}
\label{sec:mer}

Let $R\in \C$ be a simply-connected domain, bounded by piece-wise smooth curves
$\eta_1(\xi)$ and $\eta_2(\xi)$, where $x=\xi+i\eta$ is a complex number and  
$\eta_2(\xi)>\eta_1(\xi)$ for all  $\xi \in (-\infty,+\infty)$. The values 
$\eta_2=+\infty$ and $\eta_1=-\infty$ are allowed. We say that the domain
$R$ is $a$-wide on the interval $I\subset \R$ if  $\eta_2(\xi)-\eta_1(\xi)>a$ for all  
$\xi \in I$.

\begin{theorem} \label {the:1}
    A solution $f(x)$ to the equation \eqref{eq:1i} that is analytic in some
    domain $R$ that is $a$-wide on $\R$ admits a meromorphic continuation on the whole 
    complex plane and all possible singularities of $f(x)$ are first order poles.
\end{theorem}
 
\begin{proof}
Let $R_1=R$ and let $R_n$ denote the domain bounded by the curves 
$\eta_1(\xi)$ and $\eta_2(\xi)+(n-1)a$, where $n=2,3,\dots$.
We first prove that $f(x)$ can be meromorphically continued to the 
domain $R_\infty=\cup_1^\infty R_n$, i.e., to the right of
the domain $R$, by considering \eqref{eq:1i} as the Riccati equation
\begin{equation}
    \label{eq:m1}
    f'(x)=f^2(x)+h(x),
\end{equation}
where $h(x)=\m-f(x-a)'-f^2(x-a)$. The latter equation is equivalent to
to the second order linear differential equation
\begin{equation}
    \label{eq:m2}
    u''(x)+h(x)u(x)=0
\end{equation}
through the standard transformations 
\begin{equation}
    \label{eq:m3} 
u(x)=\exp{\int_{\tilde x}^x f(t)dt} \quad {\rm and} \quad f(x)=-\frac{u'(x)}{u(x)}, 
\end{equation}
where $\tilde x\in R$ is a point  such that $\tilde x-a\in R$. Note that, according to 
\eqref{eq:m3}, the function $u(x)$ is analytic in $R$.

Consider \eqref{eq:m2} in the region $Q_1$, where $Q_n=R_{n+1}\backslash  R_n$,
$n\in\Z^+$. According to the assumption of the theorem, the function $h(x)$ is analytic
in $Q_1$, so that solutions  of the linear equation \eqref{eq:m2} are analytic in $Q_1$. 
Thus, we can  analytically continue $u(x)$ onto $R_2$. If $u(x)$ does not attain
zero value in $Q_1$, we get an analytic continuation of $f(x)$ onto $R_2$ by
\eqref{eq:m3}. However, if $u(x_0)=0$ for some $x_0\in  Q_1$, then $f(x)$ has
a first order pole in $x_0$, which is an isolated singularity of $f(x)$.
Thus, we obtained the required  meromorphic continuation of $f(x)$ onto $R_2$.
This process can be continued to the domains $R_3,R_4$, etc. in the same fashion.
However, now we have to consider a possibility that $h(x)$  has a singularity
at $x_0+a$. Then, $u(x)$ and, correspondingly, $f(x)$ may have singularities at 
$x_n=x_0+na$, where $n\in\Z^+$. We need to show that these possible singularities of $f(x)$ 
are first order poles only.  

Let 
\begin{align}
    \label{eq:m3a} 
h(x)&=h_0+h_1(x-x_0)+h_2(x-x_0)^2+\dots  \quad {\rm and}  \notag \\
u(x)&=(x-x_0)+u_2(x-x_0)^2 +u_3(x-x_0)^3+\dots 
\end{align}
be the Taylor expansions of $h(x)$ and $u(x)$ near $x=x_0$ in \eqref{eq:m2}.
Comparing the like powers of $x-x_0$ in \eqref{eq:m2}, we obtain
\begin{equation}
    \label{eq:m3b} 
u_2=0,~~u_3=-\frac{h_0}{6},~~u_4=-\frac{h_1}{12},~\dots,
\end{equation}
so that the principle part of $f(x)$ at $x=x_0$ is  
$-(x-x_0)^{-1}$.

Combining the expression for $h(x)$ with \eqref{eq:m1}, we obtain  
\begin{equation}
    \label{eq:m5}
    h(x+a)=\m-h(x)-2f^2(x)=\m-h(x)-2[(\ln u(x))']^2 ~. 
\end{equation}
Direct computations show that 
\begin{equation}
    \label{eq:m5a}
  [\ln u(x)]'=\frac{1}{x-x_0}  -\frac {h_0}{3}(x-x_0)-\frac {h_1}{4}(x-x_0)^2+O(x-x_0)^3~,
\end{equation}
and
\begin{equation}
    \label{eq:m5a}
  -2([\ln u(x)]')^2=-\frac{2}{(x-x_0)^2}  +\frac {4h_0}{3}+{h_1}(x-x_0)+O(x-x_0)^2~,
\end{equation}
so that 
\begin{equation}
    \label{eq:m6}
   h_1(x+a)= h(x+a)=\frac{-2}{(x-x_0)^2}+(\m+\frac {h_0}{3})+O(x-x_0)^2~.
\end{equation}
Considering  now this equation near the point $x_1\in Q_2$,
we obtain
\begin{equation}
    \label{eq:m6a}
    h_1(x)=\frac{-2}{(x-x_1)^2}+(\m+\frac {h_0}{3})+O(x-x_1)^2~.
\end{equation}

The proof that all the points $x_n$ are either regular
points or first order poles of $f(x)$ follows by induction from the following
two lemmas.

\begin{lemma} \label {lem:1}
    The differential equation 
      \begin{equation}
      \label{eq:m7}
      u''(x)+h_n(x)u(x)=0~,
      \end{equation}
    where the coefficient $h_n(x)$ has the form
\begin{equation}
      \label{eq:m7a}
 h_n(x)= -\frac{N(N-1)}{(x-x_n)^2}+A_0+A_2(x-x_n)^2+\dots +A_{2N-2}(x-x_n)^{2N-2}
 +O(x-x_n)^{2N-1} ~,    
      \end{equation}    
posseses two linearly independent solutions 
\begin{equation}
      \label{eq:m8}
u(x)=(x-x_n)^N\left [1+u_2(x-x_n)^2+\dots+ u_{2N}(x-x_n)^{2N}+O(x-x_n)^{2N+1}\right] 
\end{equation}
 \begin{equation}
      \label{eq:m8a}     
v(x)=(x-x_n)^{-N+1}\left[1+v_2(x-x_n)^2+ \dots+ u_{2N}(x-x_n)^{2N}+O(x-x_n)^{2N+1}\right ]~ 
      \end{equation}    
that are analytic in a neighborhood of  $x=x_n$. Here $A_k,u_k,v_k\in\C$ and
$n,N\in \N$ with $n+1\geq N\geq 2$.
\end{lemma}   
    
\begin {proof}  Frobenius multipliers of   \eqref{eq:m7} at $x=x_n$ are $N$ and $-N+1$,
so the first terms of $u(x)$ and $v(x)$ are $(x-x_n)^N$ and $(x-x_n)^{-N+1}$
respectively. Suppose, 
\begin{equation}
      \label{eq:m8b}
  u(x)=(x-x_n)^{N}+\sum_{k=1}^\infty u_k (x-x_n)^{N+k}~.
\end{equation}   
Computing $u_k$, we see that the odd coefficients $u_1=u_3=
\dots=u_{2N-1}=0$, so that $u(x)$ is in the form \eqref{eq:m8}. Similar arguments
work for the second solution $v(x)$.
\end {proof}   
    
Arguments of Lemma \ref{lem:1} show that solutions to \eqref{eq:m7} have no
branching at $x=x_n$ since Frobenius multipliers are integer and there are
no logarithms. Note that the general solution to \eqref{eq:m7} has the form    
\begin{equation}
      \label{eq:m9}
  u(x)=(x-x_n)^{-N+1}\left[u_0+u_2(x-x_n)^2+\dots+ u_{2N-2}(x-x_n)^{2N-2}+
  O(x-x_n)^{2N-1}\right]
  \end{equation}    
where $u_0\not=0$. The only nontrivial special solution to  \eqref{eq:m7} is
proportional to $u(x)$ given by  \eqref{eq:m8}. In any case, the point $x_n$
is a simple pole  of the solution
$f(x)=-[\ln u(x)]'$ to \eqref{eq:m1}.
The following Lemma shows that
the new coefficient  
\begin{equation}
      \label{eq:m9a}
h_{n+1}(x)=\m-h_n(x-a)-2[(\ln u(x-a))']^2~,
\end{equation}
where $u(x)$ is a solution to \eqref{eq:m7}, has the form \eqref{eq:m7a} with another $N$.
Thus, the point $x_{n+1}$ is also a simple pole or a regular point of the solution
$f(x)$. (Note than $x_{n+1}$ is a regular point of $h_{n+1}$ if the new $N+1$.)

\begin{lemma} \label {lem:2}
      The new coefficient $h_{n+1}(x)$ has the form \eqref{eq:m7a} with the new 
      $N$ equal to $N-1$ if $u(x)$ is given by \eqref{eq:m9}  or equal to $N+1$
      if $u(x)$ is given by \eqref{eq:m8}.
\end{lemma}

\begin {proof} Consider, for example, the case when $u(x)$ is given by
\eqref{eq:m8}. Using expansion \eqref{eq:m7a} for $h_n(x)$, we obtain
the first odd coeficient $u_{2N+1}$ in \eqref{eq:m8} is
\begin{equation}
\label{eq:m10}
  u_{2N+1}=-\frac{A_{2N+1}}{4N(2N+1)}~.
\end{equation}  

On the other hand, we see that
$\ln u(x)=N\ln(x-x_n)+u_2(x-x_n)^2 + \dots$, where the leading odd
term of $\ln u(x)$ is $u_{2N+1}(x-x_n)^{2N+1}$. Then
$$
[\ln u(x)]'=(x-x_n)^{-1}\left[N+2u_2(x-x_n)^2+\dots\right]~
$$
where the leading odd term in the square brackets is 
$(2N+1)u_{2N+1}(x-x_n)^{2N+1}$. Finally, we get the leading odd term of 
$-2([\ln u(x)]')^2$ as $-4N(2N+1)u_{2N+1}(x-x_n)^{2N-1}$. So,  according to 
\eqref{eq:m9a} and \eqref{eq:m10}, the leading
odd term of $h_{n+1}(x)$ at $x_{n+1}$ is of the order $(x-x_{n+1})^{2N+1}$.
The leading term of $h_{n+1}(x)$ is $\frac{N(N-1)-2N^2}{(x-x_{n+1})^2}=
-\frac{(N+1)N}{(x-x_{n+1})^2}$. So, $h_{n+1}(x)$ has the form 
\eqref{eq:m7a} with the new $N$ equal to $N+1$. The case  when $u(x)$ is given by
\eqref{eq:m9} can be considered in a similar way.
\end {proof}

Consider, for example, the singular point $x_1$. According to \eqref{eq:m6a},
we have $N=2$ at this point. Solving the corresponding initial value problem
for \eqref{eq:m7}, we get the solution $u(x)$ of either type  \eqref{eq:m8} 
or \eqref{eq:m9}.  In any case, $f(x)$ has  a first order pole.
The corresponding function $h_2(x)$, according to Lemma \ref{lem:2},  
has a second order pole at $x=x_2$ with the principal
part $-\frac {6}{(x-x_2)^2}$ if $u(x)$ is proportional to  \eqref{eq:m8},
and is regular at $x=x_2$ if  $u(x)$ is given by  \eqref{eq:m9}.
We can continue these arguments to show that 
all points $x_n$ are either first order poles or regular points. Thus, we proved
meromorphic continuation of $f(x)$ on $R_\infty$.

To prove the meromorphic continuation to the left of the domain $R$ let us
note that the transformation $x= -t-a,~~g(t)=f(-t)$ reduces the equation \eqref{eq:1i} 
to the equation of the same type
\begin{equation}
    \label{eq:m12}
    [g(t)+g(t+a)]'+g^2(t)-g^2(t+a)=-\m,
\end{equation}
which have an analytic solution $g(t)$ on $t\in -R_\infty$. We can now use the 
previous arguments to continue $g(t)$ to the right on the whole complex plane. 
\end{proof}

\begin{corollary} \label {cor:0}
    Let $I\subset \R$ be an interval and let $f(x)$ be a
    solution to the equation \eqref{eq:1i} that is analytic in a
    domain $R$ that is $a$-wide on $I$. Then $f(x)$ admits a meromorphic continuation on  
    the strip $\Im x\in I$  and all possible singularities of $f(x)$ are first order poles.
\end{corollary}

\section{Asymptotic solutions}
\label{sec:asymp}

In this section we rewrite \eqref{eq:1i} as
\begin{equation}
    \label{eq:a1}
    \square f'(z)=\m+\D f^2(z)~,
\end{equation}
where $z=x+b,~b=\frac{a}{2}$ and the operators $\sq,\D$ act on a function $g(z)$
as 
\begin{equation}
    \label{eq:a2}
    \square g(z)=g(z+b)+g(z-b)   \and 
    \D g(z)=g(z+b)-g(z-b)~.
\end{equation}
Some important for us properties of $\sq, \D$ are given by identities
\begin{equation}
    \label{eq:a3}
    2\square [fg]=\sq f\cdot\sq g+ \D f\cdot \D g   \and 
    2\D [fg]=\D f\cdot\sq g+ \sq f\cdot \D g~, 
\end{equation}
where $f,g$ are given functions.

In this section we construct two different 
formal power series solutions $\hat f^{\pm} (z)$ to \eqref{eq:1i} and prove existence
of corresponding actual solutions $f^{\pm}_n(z)$. These solutions  are analytic in 
the corresponding sectors  $S^\pm_n, ~n=0,1,2$, specified below, when $|z|$ is 
sufficiently large and admit asymptotic expansions
\begin{equation}
    \label{eq:a4}
    f^\pm_n(z) \sim \hat f^\pm(z),~~~z\ra\infty,~z\in S^\pm_n~.
\end{equation}
This fact together 
with Theorem \ref{the:1} prove existence of non-trivial meromorphic solutions
to  \eqref{eq:1i}. 

\begin{proposition} \label {pro:3}
     Equation \eqref{eq:a1} possesses two formal power series solutions
     \begin{equation}
     \label{eq:a5}
     \hat f_\pm(z) =\pm \l\sqrt z + \frac{1}{2b}+
     \sum_{k=2}^\infty y^\pm_kz^{-\frac{k}{2}}~,
     \end{equation}
where $\l= \sqrt {-\frac{\m}{2b}}$ and the coefficients $y^\pm_k$ are defined uniquely.
\end{proposition}

\begin{proof}
The substitution $f=\l z^\hf + c+\hat y(z)$ reduces the equation \eqref{eq:a1} to
\begin{equation}
    \label{eq:a6} 
\frac{\l}{2}\sq z^{-\hf}+\sq \hat y'=\m+(\l\D z^\hf+\D\hat y)(\l\sq z^\hf+2c+\sq\hat y).
\end{equation}
Comparing leading coefficients and taking into account $\D z^\hf \sq z^\hf=2b$, 
we get $\l=\pm\sqrt {-\frac{\m}{2b}}$. Taking into account  \eqref{eq:a3}, the latter 
equation can be now rewritten as
\begin{equation}
    \label{eq:a7} 
2\l\D[ z^{\hf}\hat y]=\sq \hat y'-\D\hat y^2 -2c\D\hat y+ \l(\hf \sq z^{-\hf}-2c\D z^\hf)~.
\end{equation}
To expand $\hat y$ in powers of $z^{-\hf}$ we need the free term of  \eqref{eq:a7} to be
of order $O(z^{-1})$ or less. Thus, we obtain $c= \frac{1}{2b}$, so that the free term
is of the order  $O(z^{-\frac{3}{2}})$. Equation \eqref{eq:a7} can be now rewritten as
\begin{equation}
    \label{eq:a8} 
\D[ z^{\hf}\hat y]=\frac{1}{2\l}\left (\sq \hat y'-\D\hat y^2 -\frac{1}{b}\D\hat y\right )
+ \frac {\sq z^{-\hf}}{4}-\frac{\D z^\hf}{2b}~.
\end{equation}
It is clear that the expression in the left hand side is the dominant term of the 
latter equation and that the substitution $\hat y = 
\sum_{k=2}^\infty y^\pm_kz^{-\frac{k}{2}}$, satisfying  \eqref{eq:a8},  defines the 
coefficients $y_k$ uniquely.
\end{proof}

Let $\hat f(z)=\l\sqrt z + \frac{1}{2b}+ \dots $ denote one of the formal solutions
$\hat f^\pm(z)$. Given $\l\in\C$, we define angles 
\begin{equation}
    \label{eq:a9} 
     \b_n=\frac {2}{3}\pi(1+2n)- \frac {2}{3}\arg \l~,
\end{equation}    
where $n=0,1,2$ and $\arg\l\in [0,2\pi)$. If $\b_n+\pi$ is not a multiple of $2\pi$, 
we define the sector
$S^+_n$ on the Riemann surface of $z^{1/3}$ by extending a small sector bisected by 
$\arg z=\b_n$ independently in both positive and negative directions until it either 
hits the negative real direction 
or the ray   $\arg z=\b_n\pm\pi$ respectively. If $\b_n$ is not a multiple of $2\pi$, 
the sector $S^-_n$ on the Riemann surface of $z^{1/3}$ is defined 
by extending a small sector bisected by $\arg z=\b_n$ independently in both positive 
and negative directions until it either hits the positive real direction 
or the ray    $\arg z=\b_n\pm\pi$ respectively. In the case $\b_n=2\pi k,~k\in\Z$ or 
$\b_n=(2k+1)\pi,~k\in\Z$, the sectors $S^-_n$ or $S^+_n$ respectively are considered
to be empty. 

\begin{theorem} \label {the:2}
    If $\hat f(z)$ is a formal  solution to the equation \eqref{eq:1i} and $S^+_n$ is a 
    nonempty sector described above, then there exists and actual solution $f(z)$
    to \eqref{eq:1i} that is analytic in sufficiently remote part of any proper subsector
    $S$ of $S^+_n$ and  
\begin{equation}
    \label{eq:a10} 
     f(z) \sim \hat f(z)~,  ~~~~~~z\ra\infty,~z\in S. 
\end{equation}     
\end{theorem}

\begin {proof} Our main idea is to reduce the considered DDE to an integro-differential
equation (IDE) and to show that the latter equation can be solved by successive iterations in a
proper sectorial neighborhood of infiniti.
Substituting $f(z)=\l\sqrt z + \frac{1}{2b}+ y(z) $, we obtain equation 
\eqref{eq:a8} for $y(z)$. The inverse Laplace transform $\Lscr^{-1}$, applied  to \eqref{eq:a8},
yields (see, for example, [Ob])
\begin{align}
    \label{eq:a11} 
     -2\sinh (pb) \Lscr^{-1}[z^{\hf} y](p)&= \frac{1}{2\l}\left (-2p\cosh (pb) Y(p)+2\sinh (bp)
     \left[Y^{*2}(p)+\frac{1}{b}Y(p)\right]\right )                          \notag \\
     &+\frac{\sinh bp}{2\sqrt {\pi p}}(\coth bp-\frac {1}{bp})~,  
\end{align}  
where $Y(p)=\Lscr^{-1}[y](p)$ and $F(p)^{*2}=\int_0^pF(p-q)F(q)dq$. After a simple algebra,
the latter equation becomes 
\begin{equation}
    \label{eq:a12} 
    \Lscr^{-1}[z^{\hf} y](p)= \frac{\Xi(bp)}{2\l b}Y(p) - \frac{1}{2\l}Y^{*2}(p)
    -\frac{\Xi(bp)}{2\sqrt \pi b p^{3/2}}~,
\end{equation}
where $\Xi(x)=x\coth x-1$. Note that $\Xi(x)$ is a meromorphic function with simple poles
at the points $ {i\pi k}$, where $k\in\Z\backslash \{0\}$, and that $\Xi(x)$ has not
more than linear growth in any non-vertical direction $\arg x=const$. 

Separating the linear part $bp$ of $\Xi(bp)$ along the positive real axis, we can rewrite
\eqref{eq:a12} as
\begin{equation}
    \label{eq:a13} 
    \Lscr^{-1}[z^{\hf} y](p)-\frac{p}{2\l}Y(p)= \frac{\Xi(bp)-bp}{2\l b}Y(p) - \frac{1}{2\l}Y^{*2}(p)
    -\frac{\Xi(bp)}{2\sqrt \pi b p^{3/2}}~,
\end{equation}
where the function $\Xi(bp)-bp$ is bounded on $[0,\infty)$. Applying now the Laplace transform
$\Lscr$ to \eqref{eq:a13}, we reduce \eqref{eq:a8} to the IDE
\begin{equation}
    \label{eq:a14} 
    y'(z)+2\l z^{\hf} y(z) = \frac{1}{b}\xi(z)*y(z)-y^2(z)-\frac{\l}{b\sqrt \pi}r(z)~,
\end{equation}
where $\xi(z)=\Lscr[\Xi(bp)-bp](z)$, $r(z)=\Lscr[\Xi(bp)p^{-3/2}](z)$ and the convolution
is defined by
\begin{equation}
    \label{eq:a15} 
    f(z)*g(z)=\frac{1}{2\pi i}\int_{A-i\infty}^{A+i\infty}f(s)g(z-s)ds
    \end{equation}
with a sufficiently large $A>0$. Considering \eqref{eq:a14} as a perturbed linear 
equation $ y'(z)+2\l z^{\hf} y(z) = 0$, we rewrite the former as
    \begin{equation}
   \label{eq:a16} 
    y(z)=e^{-\frac{4}{3}\l z^{3/2}}\int_{\g(z)} e^{\frac{4}{3}\l t^{3/2}}W[y(t)]dt~,
    \end{equation}
where the nonlinear operator $W[y]$ denotes the right hand side of    \eqref{eq:a14} 
and the contour of integration $\g(z)$ is to be specified below. 
Equation \eqref{eq:a16} can be rewritten in the operator form as $y=\Iscr W[y]$. 

Let us assume for a while that $\frac{\pi}{4}<\arg \l < \frac{7\pi}{4}$
and consider sector $S^+_0$. According to \eqref{eq:a9}, this choice of $\l$ allows us 
to find a  proper closed subsector $S\subset S^+_0$ that contains the right half-plane. 
We want to solve the IDE \eqref{eq:a14}  
in a sufficiently remote part of the sector $S$ by successive approximations. 
In order to formulate the statement more precisely we need to introduce
the following notations.

Let  $\Sigma$ be the image of the sector  $S$ under
the transformation $\z=z^{3/2}$.
Let $z_0$ be a sufficiently remote point on the ray $\arg z=\b_0$ and 
let $\Sigma_{\z_0}$, where $\z_0=z_0^{3/2}$  denote the parallel shift of $\Sigma$ so 
that the vertex of $\Sigma$ is shifted to $\z_0$. For every $\z\in\Sigma_{\z_0}$ we define 
a contour $\G(\z)$ as a ray eminating from $\z$ and such that: $\G(\z)=\{\t~:~\arg (\t-\z)
=\frac{3}{2}\b_0\}$ if $|\arg \z-\frac{3}{2}\b_0|<\frac {\pi}{2}$; 
$\G(\z)=\{\t~:~\arg (\t-\z)
=\arg\z +\frac{\pi}{2}\}$ if $\arg \z-\frac{3}{2}\b_0<-\frac {\pi}{2}$ and; 
$\G(\z)=\{\t~:~\arg (\t-\z)
=\arg \z-\frac{\pi}{2}\}$ if $\arg \z-\frac{3}{2}\b_0>\frac {\pi}{2}$. Then the region
$S_{z_0}$  and the contour $\g(z)$ are the images of $\Sigma_{\z_0}$ 
and $\G(\z)$ under the transformation $z=\z^{2/3}$.

Let $y_0(z)\equiv 0$, $y_k(z)=\Iscr\circ W[y_{k-1}](z),~k=1,2,\dots$ and $\d y_k=
y_k-y_{k-1}$. We will show that the solution to \eqref{eq:a16} is given by
 \begin{equation}
   \label{eq:a17} 
    y(z)=\sum_{k=1}^\infty \d y_k~,
    \end{equation}
where the series converges absolutely and uniformly in $S_{z_0}$ for sufficiently
large $|z_0|$. This can be done by introducing the Banach space $\Bscr$ of functions $h(z)$,
such that $h$ is analytic on  $S_{z_0}$ and  satisfy $| h(z) | \leq B |z^{-2}|$ there with 
some constant $B>0$.  (Note that $B$ depends on $h$.) The norm of  $h\in\Bscr$
is $\sup_{z\in S_{z_0}} | z^{-2}h(z) |$. According to Lemma 14.2 from \cite{Wa},
the integral operator $\Iscr:~\Bscr \ra\Bscr$ is a bounded linear operator,
where $\V\Iscr\V$ is proportional to $|z_0^{-\hf}|$.
We start to study the nonlinear operator $\Wscr[y]$ by considering  the convolution 

\begin{align}
    \label{eq:a18} 
    \xi(z)*y(z)&=\Lscr[(\Xi(bp)-bp)Y(p)](z)=\Lscr[\{bp(\coth bp-1)-e^{-2bp}\}Y(p)](z)  \notag \\
    &+\Lscr[(e^{-2bp}-1)Y(p)](z)=\mu(z)*y(z)+y(z+2b)-y(z)~,
    \end{align}
where, according to \cite{OB}, 
\begin{equation}
    \label{eq:a18a} 
    \mu(z)=\Lscr[\{bp(\coth bp-1)-e^{-2bp}]=\frac{1}{2b}\Psi'(1+\frac{z}{2b})-\frac{1}{z+2b}~.
    \end{equation}
Here $\Psi(x)=\frac{\G'(x)}{\G(x)}$ denotes the logarithmic derivative of the Gamma-function.
It is clear that $\m(z)$ is a meromorphic function with double poles at $z=-2kb, ~k\in\Z^+$.
The asymptotic expansion 
\begin{equation}
    \label{eq:a19} 
    \Psi'(x) ~\sim ~ \frac {1}{z}+\frac {1}{2z^2}+\sum_{k=1}^\infty\frac {B_{2k}}{z^{2k+1}},
    ~~~~z\ra\infty,~~|\arg z |<\pi~,
    \end{equation}
where $B_{2k}$ are Bernoulli numbers, follows from the Stirling formula  (see \cite {GR}).
Combining the latter  facts, we obtain the estimate
\begin{equation}
    \label{eq:a20} 
    |\m(z)|\leq \frac {b\Psi_0}{|z+2b|^2},~~~~~~z\in S~,
    \end{equation}
where the constant $\Psi_0$ depends only on the sector $S$. 

\begin{proposition} \label {pro:4}
    If $y\in\Bscr$ then $\m(z)*y(z) \in\Bscr$ and
     \begin{equation}
     \label{eq:a21}
     \V \frac {1}{b}\m*y\V\leq M\|y\|~,
     \end{equation}
   where $M>0$ does not depend on $y$ and $z_0$.  
\end{proposition}

\begin {proof} Consider first the case $z$ belongs to the right half-plane 
$\hat S=\{z~:~\Re z \geq mz_0\}$, where the number $m>0$ is choosen 
so that $\hat S\subset S_{z_0}$. Note that $m$ does not depend on $|z_0|$.
Setting 
$A=mz_0$ in \eqref{eq:a15}, we obtain
    \begin{equation}
    \label{eq:a22} 
    \m(z)*y(z)=\frac{1}{2\pi i}\int_{A-i\infty}^{A+i\infty}y(s)\m(z-s)ds
    \end{equation}
Note that we are integrating over the boundary of $\hat S$  and thus $\Re (z-s)\geq 0$. 
The fact that $\m(z)*y(z)$ is analytic in $\hat S$
follows from the properties of $\m(z)$ and $y(z)$ immediately. Utilizing \eqref{eq:a20} and 
the fact that $y\in\Bscr$, we obtain
   \begin{align}
    \label{eq:a23} 
   \frac{1}{b\|y\|}| \m(z)*y(z)|&\leq \frac{\Psi_0}{2\pi }
   \int_{A-i\infty}^{A+i\infty}\frac {|ds|}{|s|^2|z+2b-s|^2}  \notag \\
   &=\frac{\Psi_0}{2\pi }
   \int_{-\infty}^{+\infty}\frac {d\eta}{(A^2+\eta^2)[(u-A)^2+(v-\eta)^2]}~,
    \end{align}
where $s=A+i\eta$ and $z+2b=u+iv$ with $\eta,u,v\in\R$. 

Let $I$ be the latter integral and $g(\eta)$ denote its integrand, which has simple
poles at the points $\eta=iA$ and $\eta=v+i(u-A)$ in the upper half-plane. Computing
$I$ via the residues of $g(\eta)$ in the upper half-plane, we obtain
 \begin{align}
    \label{eq:a24} 
    I=2\pi i\sum\Res g(\eta)&=\frac{\pi}{v+i(u-2a)}\left[\frac {1}{A(v-iu)}+
    \frac {1}{(u-A)(v+iu)}\right]                           \notag \\
    &=\frac {\pi }{A|z+2b|^2}\left (1+\frac{A}{u-A}\right)
    \end{align}
Thus
    \begin{equation}
    \label{eq:a25} 
   \frac{1}{b\|y\|}| \m(z)*y(z)|\leq\frac{\Psi_0}{2b |z|^2}~
    \end{equation}
since $A$ can be taken greater than $2b$.

Consider now the half-plane $\hat S_\phi$ that is obtained   by rotating $\hat S$
on the angle $\phi$, where $\phi$ is choosen in such a way that $\hat S_\phi\subset S_{z_0}$.
Define now another convolution $m*_\phi y$ by \eqref{eq:a22}, where the contour of
integration is the boundary of $\hat S_\phi$. Using the same arguments as above, we
can obtain the estimate
    \begin{equation}
    \label{eq:a26} 
   \frac{1}{b\|y\|}| \m(z)*_\phi y(z)|\leq\frac{M_\phi}{ |z|^2}~,
    \end{equation}
where $z\in\hat S_\phi$ and $M_\phi$ continuously depends on $\phi$. However,
$m(z)*_\phi y(z)$ is an analytic continuation of $m(z)*y(z)$, since the functions 
coincide on $\hat S_\phi\cap \hat S$. Taking $M=\max M_\phi$, where $M_0=\frac{\Psi_0}{2b}$,
we complete the proof of the proposition.
\end{proof}

To complete the proof of the theorem we use the standard technique to show the
convergence of iterations \eqref{eq:a17}. Using properties of the Laplace transform
and of the operator $\Iscr$, one can show that $y_1\in\Bscr$. Let $K=\|y_1\|$, and 
let us prove by induction that $\|\d y_n\|\leq 2^{1-n}K$ if $|z_0|$ is sufficiently large.
Indeed, according  to the estimate of the convolution,
   \begin{equation}
    \label{eq:a27} 
    \|\d y_n\|\leq \|\Iscr\|\left (M+2+\frac{4K}{|z|^2}\right)\|\d y_{n-1}\|~,
    \end{equation}
where  the induction assumption    
\begin{equation}
    \label{eq:a28} 
    \| y_{n-1}^2-y_{n-2}^2\|\leq \| y_{n-1}+y_{n_2}\|\frac {|\d y_{n-1}\|}{|z|^2}
    \leq 4K\frac {|\d y_{n-1}\|}{|z|^2}~
    \end{equation}
was  used to estimate the nonlinear term of $\Wscr$.
It remains to choose $z_0$ so that  $\|\Iscr\|\left (M+2+\frac{4K}{|z|^2}\right)<\hf$
for $z\in S_{z_0}$ to complete the proof of the theorem for  $\frac{\pi}{4}<\arg \l < 
\frac{7\pi}{4}$ the sector $S^+_0$. 

Let us now consider the general case $0\leq \arg \l<2\pi$. The sector  $S^+_0$
is given by $-\frac{\pi+2\arg\l}{3}<\arg z<\pi$ if $\arg \l\in[0,\pi]$ and 
by $-\pi <\arg z< \frac{\pi-2\arg\l}{3}$ if $\arg \l\in[\pi,2\pi]$. Note that 
the opening of the  sector $S^+_0$ is greater than $\pi$ for any $\arg \l$ and that  $S^+_0$
contains the right hlaf-plane if $\frac{\pi}{4}<\arg \l < \frac{7\pi}{4}$.
Let us now choose  a  proper closed subsector $S\subset S^+_0$ of opening 
greater than $\pi$, and let $\arg z=\a$ be the bisector of $S$.Clearly,
$|\a|<\frac{\pi}{2}$. Let 
\begin{equation}
    \label{eq:a29} 
    \Lscr_\a [Y](z)=\int_0^{e^{-i\a}\infty} e^{-zp}Y(p)dp
    \end{equation}
define the Laplace transform along the ray $\arg p=-\a$. The contour for the
corresponding inverse Laplace transform as well as for the corresponding
convolution $*_\a$ is a straight line perpendicular to $\arg z=\a$.
Therefore, we can use our previous arguments to show the uniform and absolute
convergence of iterations  \eqref{eq:a17} in a properly constructed $S_{z_0}
\subset S$. In the same fashion the theorem can be proven for any nonempty
sector $S^+_n$, $n\in\Z$.

Recall that the function $\Xi(bp)$ has poles on the imaginary axis and, therefore,
$\Lscr_\a \Xi$ is not defined for $\a=\pm\frac{\pi}{2}$. However, we can define
$\Lscr_\a \Xi$ for $\a$ such that $|\a-\pi|< \frac{\pi}{2}$ and repeat the 
previous arguments for sectors $S^-_n$, $n\in\Z$.

\end{proof}

\begin{corollary} \label {cor:1}
     
Let $f^\pm_n, ~n=0,1,2$ denote a solution of \eqref{eq:1i}, analytic in some remote
part of $S^\pm_n$ as described in \ref {the:2}. Then, according to \ref {the:1},
$f^\pm_n$ is a meromorphic function on $\C$ that can have only simple poles.

\end{corollary}

\end{document}